\documentclass[aps,showpacs]{revtex4}

\usepackage{graphicx}
\usepackage{amsmath}
\usepackage{amssymb}

\begin{document}

\title{Constraining Ho\v{r}ava-Lifshitz gravity by weak and strong
gravitational lensing}
\author{Zsolt Horv\'{a}th}
\email{zshorvath@titan.physx.u-szeged.hu}
\affiliation{Departments of Theoretical and Experimental Physics, University of Szeged, D%
\'{o}m t\'{e}r 9, Szeged 6720, Hungary}
\author{L\'{a}szl\'{o} \'{A}. Gergely}
\email{gergely@physx.u-szeged.hu}
\affiliation{Departments of Theoretical and Experimental Physics, University of Szeged, D%
\'{o}m t\'{e}r 9, Szeged 6720, Hungary}
\author{Zolt\'{a}n Keresztes}
\email{zkeresztes@titan.physx.u-szeged.hu}
\affiliation{Departments of Theoretical and Experimental Physics, University of Szeged, D%
\'{o}m t\'{e}r 9, Szeged 6720, Hungary}
\author{Tiberiu Harko}
\email{harko@hkucc.hku.hk}
\affiliation{Department of Physics and Center for Theoretical and Computational Physics,
The University of Hong Kong, Pok Fu Lam Road, Hong Kong}
\author{Francisco S. N. Lobo}
\email{flobo@cii.fc.ul.pt}
\affiliation{Centro de Astronomia e Astrof\'{\i}sica da Universidade de Lisboa, Campo
Grande, Edif\'{\i}cio C8 1749-016 Lisboa, Portugal}

\begin{abstract}
We discuss gravitational lensing in the Kehagias-Sfetsos space-time emerging
in the framework of Ho\v{r}ava-Lifshitz gravity. In weak lensing we show
that there are three regimes, depending on the value of $\bar{\lambda}%
=1/\omega d^{2}$, where $\omega $ is the Ho\v{r}ava-Lifshitz parameter and $%
d $ characterizes the lensing geometry. When $\bar{\lambda}$ is close to
zero, light deflection typically produces two images, as in Schwarzschild
lensing. For very large $\bar{\lambda}$ the space-time approaches flatness,
therefore there is only one undeflected image. In the intermediate range of $%
\bar{\lambda}$ only the upper focused image is produced due to the existence
of a maximal deflection angle $\delta _\text{max}$, a feature inexistent in
the Schwarzschild weak lensing. We also discuss the location of Einstein
rings, and determine the range of the Ho\v{r}ava-Lifshitz parameter
compatible with present day lensing observations. Finally, we analyze in the
strong lensing regime the first two relativistic Einstein rings and
determine the constraints on the parameter range to be imposed by
forthcoming experiments.
\end{abstract}

\pacs{04.50.Kd, 95.30.Sf, 98.35.Jk, 98.62.Sb}
\maketitle

\section{Introduction}

Recently, Ho\v{r}ava proposed a renormalizable field theoretical model which
can be interpreted as a complete theory of gravity \cite%
{Horava:2008ih,Horava:2009uw}. In the infrared (IR) energy scales, the
theory reduces to Einstein gravity with a nonvanishing cosmological
constant. However, in the ultraviolet (UV) energy scales, the theory
exhibits an anisotropic Lifshitz scaling between time and space given by $%
x^{i}\rightarrow lx^{i}$ and $t\rightarrow l^{z}t$, where $z$ is the scaling
exponent. Due to the latter anisotropic scaling the model is denoted Ho\v{r}%
ava-Lifshitz gravity in the literature. Taking into account these novel
features, Ho\v{r}ava-Lifshitz gravity has received a tremendous amount of
attention. As the literature is rather extensive, we refer the reader to 
\cite{Visser:2011mf} for a recent status of the theory. In addition to the
formal issues, applications have been extensively explored, ranging from
cosmology, dark energy, dark matter to spherically symmetric or rotating
solutions.

In fact, several versions of Ho\v{r}ava gravity have been proposed in the
literature \cite{Visser:2011mf}. The relevant version for cosmology was the
introduction of an IR modification term containing an arbitrary cosmological
constant, representing the analogs of the standard Schwarzschild-(anti) de
Sitter solutions, which were absent in the original Ho\v{r}ava model. In
this context, IR-modified Ho\v{r}ava gravity seems to be consistent with the
current observational data \cite{observ,Harko:2009qr}, but in order to test
its viability more observational constraints are necessary.

In this work, we discuss the position of images formed in gravitational
lensing by the Kehagias-Sfetsos asymptotically flat space-time \cite%
{Kehagias:2009is} in the framework of Ho\v{r}ava-Lifshitz gravity. Note that
gravitational lensing has become a useful tool in measuring certain
properties of gravitational fields and it has now been employed to study the
large scale structure of the Universe, to determine behavior of compact
stellar objects and to search for dark matter candidates. In what follows we
advocate the idea that gravitational lensing might also be used to
discriminate which of the various gravitational theories is correct. Indeed,
this approach was followed in Ref. \cite{HG} in the context of 5-dimensional
brane-world theories. In particular, by computing the bending angles and
image brightness changes that occur due to the passage of photons past
lensing object were determined to distinguish a general relativistic black
hole from the black holes predicted by an alternative theory.

We follow the approach proposed in \cite{VirEllis}, where under some
physically realistic assumptions, a simple lens equation was obtained that
allows arbitrary small as well as large light deflection angles. This lens
equation has been widely used in the literature for studying strong field
gravitational lensing. In fact, in the last few years there has been a
growing interest in studying weak as well as strong field lensing by a wide
plethora of compact objects (we refer the reader to Ref. \cite%
{Virbhadra:2008ws} and references therein and to Refs. \cite{stronglens} for
pioneering contributions on strong gravitational lensing).

This paper is organized as follows. In Sec. \ref{secIII}, we present the
basics of weak lensing in a generic static and spherically symmetric
space-time and introduce the relevant notations. In Sec. \ref{secIV}, the
image locations in the Kehagias-Sfetsos space-time are analyzed, in
particular, the formation of Einstein rings. In Sec. \ref{secV} the weak
lensing properties, and in Sec. \ref{secVI} the strong lensing properties of
the Kehagias-Sfetsos space-time are analyzed. Finally, in Sec. \ref%
{sec:concl} we present our conclusions.

\section{Weak lensing in Kehagias-Sfetsos geometry \label{secIII}}

\begin{figure}[h]
\begin{center}
\includegraphics[width=10cm]{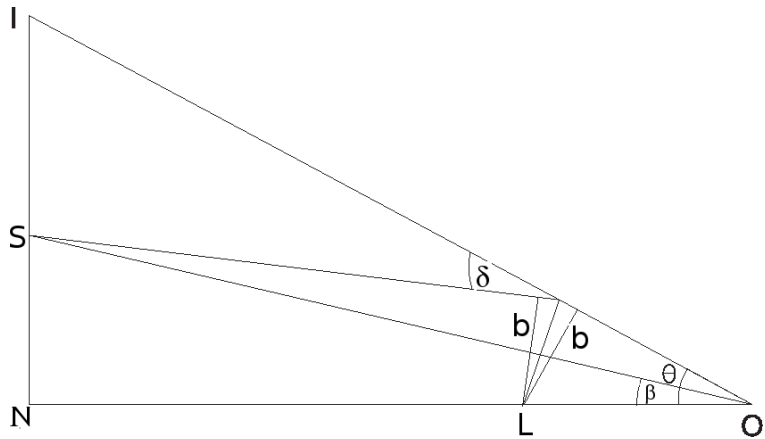}
\end{center}
\caption{The lensing geometry for positive apparent angle $ \theta $
and deflection angle $ \delta $.}
\label{fig1}
\end{figure}

In this section we summarize the basics of weak lensing in a generic static
and spherically symmetric space-time and introduce the relevant notations.
The optical axis is defined by the lensing object ($L$) and the observer ($O$
) (Fig \ref{fig1}). Relative to this axis and seen by $O$, the source ($S$)
is under angle $\beta $, chosen positive by convention ($S$ is always on the
upper part of the optical axis). Due to the lensing effect the image of the
source ($I$) appears shifted away. The angle $\theta =\widehat{IOL}$
characterizes the apparent position of the source and it is either positive
(for images on the upper side of the optical axis) or negative (for images
below the optical axis). Let us denote $s=$sgn $\theta $. Finally, the
deflection angle $\delta =\widehat{SAI}$ shows the change in the direction
of light as compared to an undeflected trajectory. We follow the convention
of $\delta >0$ whenever the light is bent towards the optical axis and $%
\delta <0$ otherwise. The projection of both $S$ and $I$ onto the optical
axis ($N$) lies at distance $D_{LS}$ from the lensing object and at $D_{S}$
from the observer. The observer-lensing object distance therefore is $%
D_{L}=D_{S}-D_{LS}$. We denote $d=D_{L}D_{S}/D_{LS}$. The impact parameter
is $b=D_{l}\sin \left( s\theta \right) $.

The deflection angle $\delta $ can be calculated by comparing the two
asymptotic behaviors of the null geodesics in the $\Theta =\pi /2$\ plane,
and it is given by the elliptic integral \cite{WEINBERG}: 
\begin{equation}
\delta \left( r_\text{min}\right) =2\int_{r_\text{min}}^{\infty }\frac{1}{r}%
\left[ \frac{g_{rr}\left( r\right) }{\frac{g_{tt}\left( r_\text{min}\right) 
}{g_{tt}\left( r\right) }\left( \frac{r}{r_\text{min}}\right) ^{2}-1}\right]
^{1/2}dr-\pi ~, \label{defl}
\end{equation}%
where $r_\text{min}$ is the distance of closest approach to the lens: 
\begin{equation}
\frac{dr}{d\varphi }\left( r_\text{min}\right) =0~. \label{rela0}
\end{equation}

\subsection{The Kehagias-Sfetsos geometry}

In the framework of Ho\v{r}ava-Lifshitz gravity the geometry of the
Kehagias-Sfetsos asymptotically flat space-time \cite{Kehagias:2009is} is
given by the following static and spherically symmetric solution 
\begin{equation}
ds^{2}=g_{tt}\left( r\right) dt^{2}+g_{rr}(r)dr^{2}+r^{2}\left( d\Theta
^{2}+\sin ^{2}\Theta d\varphi ^{2}\right) \,, \label{metric}
\end{equation}%
where the metric functions are provided by 
\begin{equation}
-g_{tt}(r)=1/g_{rr}(r)=1+\omega r^{2}\left[ 1-\left( 1+\frac{4GM}{%
c^{2}\omega r^{3}}\right) ^{1/2}\right] \,, \label{ks}
\end{equation}%
$c$ is the speed of light, $G$ is Newton's constant, $M$ is the total mass
of the black hole, and $\omega $ is the Ho\v{r}ava-Lifshitz parameter (for
details we refer the reader to \cite{Kehagias:2009is}).

The Kehagias-Sfetsos space-time (\ref{ks}) presents a curvature singularity
in the origin. The quantity $\omega _{0}=m^{2}\omega $ ($m\equiv MG/c^{2}$)$%
\ $was found useful in \cite{Harko:2009qr} for confronting the space-time
with Solar System tests. There are two additional coordinate singularities at%
\begin{equation}
r_{\pm }=m\left( 1\pm \sqrt{1-\frac{1}{2\omega _{0}}}\right) ~, \label{hors}
\end{equation}%
however they can be transformed away by introducing Eddington-Finkelstein
type coordinates ($v,r$). The apparent horizon at $r_{+}$, defined by $%
dr/dv=0$ (the outer boundary of the trapped region) represents an event
horizon, a property which holds in general for stationary space-times. In
the limit $\omega _{0}\rightarrow \infty $ the outer horizon $r_{+}$
approaches the Schwarzschild horizon $2m$ and the inner horizon $r_{-}$
approaches the central singularity. Unless $\omega _{0}\geq \omega
_{0}^{extr}=1/2$, the solutions $r_{\pm }$ become imaginary, the horizon is
absent and the singularity at the origin becomes naked. This does not really
represent a restriction on the range of $\omega _{0}$, as one can always
match the Kehagias-Sfetsos solution with a suitable interior stellar
solution at some surface $R_{star}\gg r_{+}$, similarly as when black hole
solutions are employed to describe stellar exteriors.

We note additional difficulties related to the black hole interpretation due
to the nonrelativistic dispersion relations, see Ref. \cite{Wang} with the
relevant references and discussion presented there. However, for spherically
symmetric solutions of the infrared limit of Ho\v{r}ava-Lifshitz gravity
inside the metric horizon there is a universal horizon even in the presence
of arbitrarily high propagation speeds \cite{Jacobson}. Ref. \cite{nullgeod}
presents an action principle for Ho\v{r}ava-Lifshitz gravity, based on a
Foliation Preserving Diffeomorphisms, according to which massive particles
do not follow geodesic paths, however massless particles follow null
geodesics. Therefore the expression of the deflection angle (\ref{defl}) can
be applied for the light propagating in the Kehagias-Sfetsos space-time (\ref%
{ks}).

\subsection{Deflection angle and lens equation}

By performing a change of variable $\alpha =\arcsin (r_\text{min}/r)$, Eq. (%
\ref{defl}) gives the deflection angle

\begin{equation}
\delta \left( x_{0}\right) =2\int_{0}^{\pi /2}\left[ 1+\frac{8\left( \sin
^{3}\alpha -1\right) \left( \tan ^{2}\alpha +1\right) \omega _{0}x_{0}}{%
\left( 16\omega _{0}^{2}x_{0}^{4}+8\omega _{0}x_{0}\sin ^{3}\alpha \right)
^{1/2}+\left( 16\omega _{0}^{2}x_{0}^{4}+8\omega _{0}x_{0}\right) ^{1/2}}%
\right] ^{-1/2}d\alpha -\pi ~. \label{deltax0}
\end{equation}%
When $\omega _{0}\rightarrow \infty $ we obtain the Schwarzschild limit of
the deflection angle, increasing with $x_{0}=r_\text{min}/2m$ (the distance
of closest approach to the lensing object). By contrast, the limit $\omega
_{0}\rightarrow 0$ gives a flat space-time and a vanishing deflection angle.
With no lens mass there is also no deflection, irrespective of the value of $%
\omega $, as $\omega _{0}$ vanishes. The quantities $m$, $x_{0\text{ }}$and $%
\delta \left( x_{0}\right) $ have all the same sign, as in Schwarzschild
lensing.

Next we replace the condition (\ref{rela0}) with an algebraic equation,
following the logic of Ref. \cite{GERGELY}. Light-like motions in the
equatorial plane of the metric (\ref{metric}) are governed by the Lagrangian 
\begin{equation}
2\mathcal{L}=g_{tt}\left( r\right) \dot{t}^{2}+g_{rr}(r)\dot{r}^{2}+r^{2}%
\dot{\varphi}^{2}=0~, \label{Lag}
\end{equation}%
the dot representing a derivative with respect to a parameter of the curve.
The cyclic coordinates $t$ and $\varphi $ imply two constants of motion $%
E=-g_{tt}\dot{t}$ and $L=r^{2}\dot{\varphi}$. Reinserting these in Eq. (\ref%
{Lag}) we obtain in terms of $u=r^{-1}$ the equation for the trajectory 
\begin{equation}
g_{rr}u^{\prime }=g_{tt}^{-1}\left( \frac{E}{L}\right) ^{2}-u^{2}~,
\label{trajectory}
\end{equation}%
where a prime denotes the derivative with respect to $\varphi $. The
condition (\ref{rela0}) gives 
\begin{equation}
0=\left( \frac{L}{E}\right) ^{2}g_{tt}\left( r_\text{min}\right) +r_\text{min%
}^{2}~.
\end{equation}%
To zeroth order in the lensing the trajectory would be a straight line with $%
\left. r_\text{min}\right\vert _{\text{leading order}}\equiv L/E=b=D_{l}\sin
\left( s\theta \right) $. Therefore we could replace the differential
condition (\ref{rela0}) with the algebraic relation 
\begin{equation}
0=g_{tt}\left( r_\text{min}\right) D_{L}^{2}\sin ^{2}\theta +r_\text{min}%
^{2}~. \label{rela}
\end{equation}%
The apparent angle $\theta $, under which an image appears, is found from
the Virbhadra-Ellis lens equation \cite{Bozza,Virbhadra:2008ws}: 
\begin{equation}
0=\tan \theta -\tan \beta -\frac{D_{LS}}{D_{S}}\left[ \tan \theta +\tan
\left( s\delta -\theta \right) \right] ~. \label{lens_Ellis}
\end{equation}%
The numerical solution of Eqs. (\ref{deltax0}), (\ref{rela}) and (\ref%
{lens_Ellis}), together with the expressions of $g_{tt}$ and $g_{rr}$ gives
the loci of the images.

The simplest example for solving the system above is for a Schwarzschild
black hole $-g_{tt}=1-2m/r$. After expanding the integrand in Eq. (\ref{defl}%
), and Eq. (\ref{rela}) in powers of $\bar{\varepsilon}$, to\textit{\
leading order} we get $\delta =4m/r_\text{min}$ and $\theta =sr_\text{min}%
/D_{L}$, which together give $\delta =4m/D_{L}\left\vert \theta \right\vert $%
. The positions of the images are 
\begin{equation}
\theta _{1,2}=\frac{\beta }{2}\pm \sqrt{\frac{\beta ^{2}}{4}+4\bar{%
\varepsilon}}~. \label{solfor}
\end{equation}%
In what follows, we will need the position of images for $g_{tt}$ of the
Kehagias-Sfetsos space-time.

\subsection{Image locations in the Kehagias-Sfetsos space-time \label{secIV}}

We first define $\bar{\varepsilon}=m/d$, then by introducing normalized
parameters $\theta /\beta $ and $\bar{\varepsilon}/\beta ^{2}$ provided that 
$\beta \neq 0$, the dependence of the apparent angles from $\beta $ is
eliminated.$\ $ As for the mass, we also define a useful parameter $\bar{%
\lambda}=1/\omega d^{2}$, encompassing information about the Ho\v{r}%
ava-Lifshitz parameter $\omega $ and the lensing geometry. The parameters $%
\bar{\varepsilon}$, $\bar{\lambda}$ and $\omega _{0}$ obey%
\begin{equation}
\omega _{0}\bar{\lambda}=\bar{\varepsilon}^{2}~. \label{params}
\end{equation}%
Then this leads to the black hole condition 
\begin{equation}
\bar{\varepsilon}^{2}\geq \frac{\bar{\lambda}}{2}~. \label{BHcond}
\end{equation}

The weak gravity sector of the Kehagias-Sfetsos space-time occurs when $%
-g_{tt}-1\ll 1$. This can either occur close to the Schwarzschild limit ($%
\omega \rightarrow \infty $) or when $\omega \rightarrow 0$ (thus $\bar{%
\lambda}\rightarrow \infty $). The latter condition renders the geometry
outside the black hole parameter region (large $\bar{\lambda}$ and finite $m$%
, thus Eq. (\ref{BHcond}) is not obeyed). In this limit the Kehagias-Sfetsos
space-time becomes flat without approximating Schwarzschild, but rather a
naked singularity (unless matched with a stellar solution replacing the
central region).

Negative mass parameters are allowed only for $1+4m/\omega r^{3}\geq 0$,
thus for 
\begin{equation}
\bar{\lambda}\leq \bar{\lambda}^\text{crit}=\frac{1}{4}\left( \frac{r_\text{%
min}}{d}\right) ^{3}\left( -\bar{\varepsilon}\right) ^{-1}~. \label{cond}
\end{equation}

\subsubsection{Einstein rings}

\begin{figure}[tbp]
\begin{center}
\includegraphics[height=7cm]{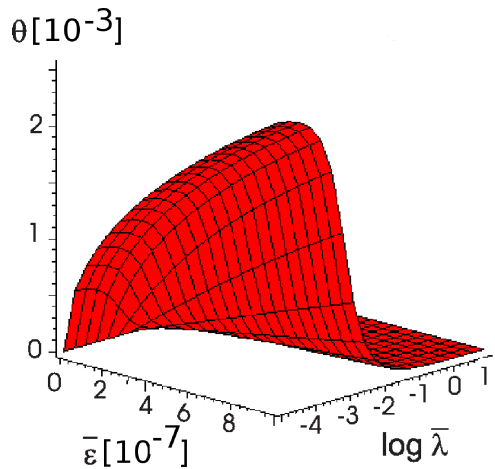}
\end{center}
\caption{The radius of the Einstein ring as function of the mass parameter $%
\bar{ \varepsilon}$ and Ho\v{r}ava-Lifshitz parameter $\bar{ %
\lambda}$ for the Kehagias-Sfetsos space-time, assuming $D_{LS}/D_{L}=2$.
For $\bar{ \lambda}\rightarrow 0$ we reobtain the Schwarzschild
result. With increasing $\bar{ \lambda}$, as the metric approaches
flatness, the radius of the Einstein ring shrinks, and tends to zero for
large values of $\bar{ \lambda}$.}
\label{fig2}
\end{figure}

The Einstein ring with radius $\theta _{E}\geq 0$ occurs when the source,
lens and observer are on the same axis. The numerical solution of Eqs.~(\ref%
{deltax0}), (\ref{rela}) and (\ref{lens_Ellis}) with the conditions $\beta
=0 $ and $s=1$ gives $\left( \theta _{E},r_\text{min}/d,\delta \right) $ as
functions of the parameters $\bar{\lambda}$ and $\bar{\varepsilon}$. The
radius of the ring is represented on Fig.~\ref{fig2} for the particular
configuration $D_{LS}/D_{L}=2$. For $\bar{\lambda}\rightarrow 0$ we get the
Schwarzschild lensing (the half-parabola shaped section of the surface with
the largest opening with respect to the $\theta =0$ plane). As $\bar{\lambda}
$ increases, the opening of the half-parabola decreases, indicating the
weakening of gravity (for a given $\bar{\varepsilon}$). At sufficiently
large $\bar{\lambda}$ the contribution of $\bar{\varepsilon}$ becomes
irrelevant and the geometry flattens, rendering the radius of the Einstein
ring $\theta _{E}\rightarrow 0$. For a given mass and lensing geometry, the
Einstein angles in the Kehagias-Sfetsos space-time are always smaller than
their Schwarzschild value.

\subsubsection{Images}

\begin{figure}[tbp]
\includegraphics[width=18cm]{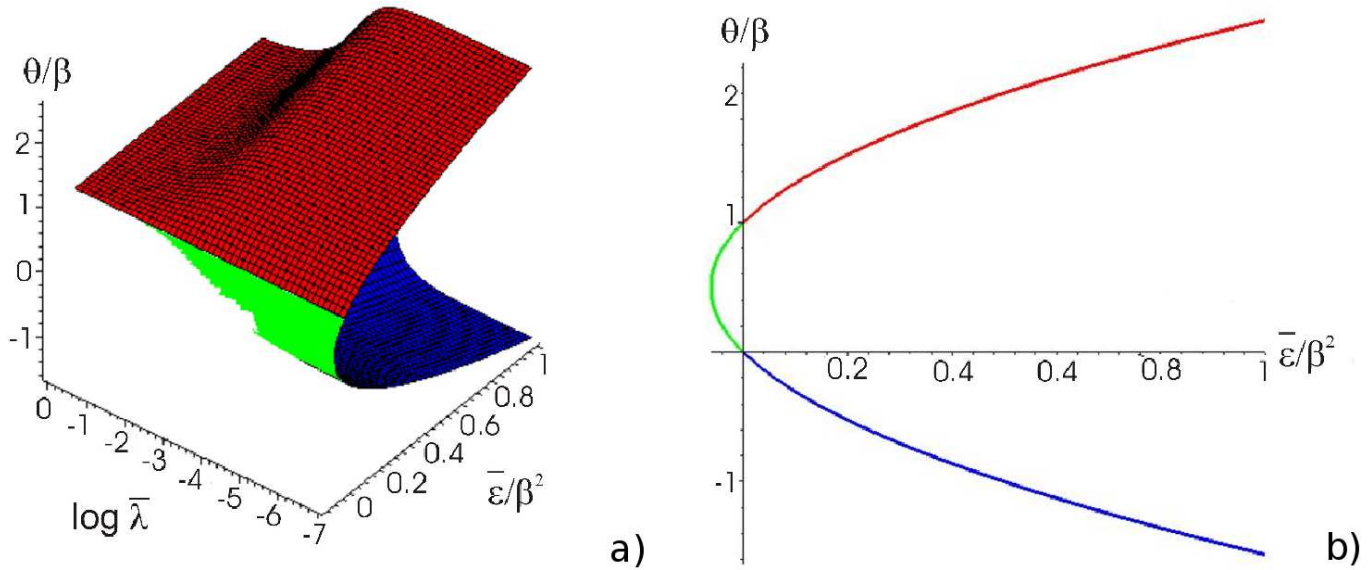}
\caption{The position of the images as function of the mass parameter $\bar{%
 \varepsilon}$ and Ho\v{r}ava-Lifshitz parameter $\log \bar{ %
\lambda}$ for the Kehagias-Sfetsos space-time (with $D_{LS}/D_{L}=2$ and\ $%
 \beta =10^{-3}$ rad) is represented on panel a). The three surfaces
on the figure refer to: (1) the focused positive image is represented by the
upper (red) surface; (2) the focused negative image is seen underneath
(blue); (3) the scattered images (with $0< \theta < \beta $)
are found in the junction of the two surfaces mentioned generated by a
negative mass (green). For $\bar{ \lambda}\rightarrow 0$ we reobtain
the Schwarzschild result, the parabola shown on panel b). With increasing $%
\bar{ \lambda}$ only the positive focused image is left. For very
large $\bar{ \lambda}$, the metric approaches flatness.}
\label{fig3}
\end{figure}
\begin{figure}[tbp]
\includegraphics[height=7cm]{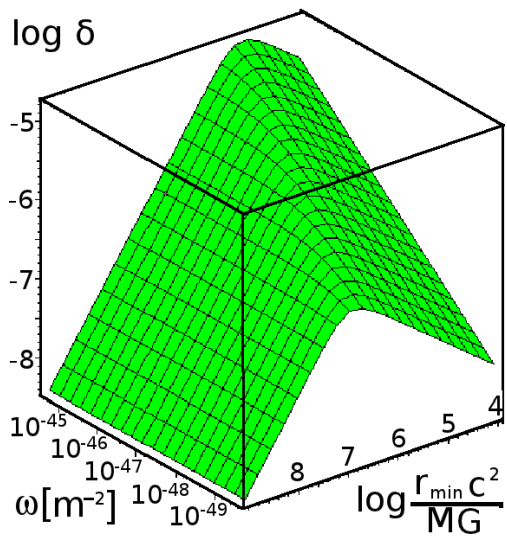}
\caption{The logarithm of the deflection angle $ \delta $ as function
of the distance of minimal appoach (represented in units $m$, on logarithmic
scale) and Ho\v{r}ava-Lifshitz parameter $ \omega $. The represented
range of $ \omega $ corresponds to the range of $\bar{ \lambda}
$ of Fig \ref{fig3}, when the lens mass is $m=4.284\cdot 10^{14}$
meter, $D_{S}=4.190\cdot 10^{25}$ meter and $D_{LS}/D_{L}=2$. For every $%
 \omega $ there is a maximal deflection angle $ \delta _{\text{%
max}}\ $, corresponding to certain $r_{\text{crit}}$. \ The critical $r_{%
\text{crit}}~$distance decreases with increasing $ \omega $. (Near
the Schwarzschild limit $r_{\text{crit}}$ shelters below the horizon and
tends to 0 when $ \omega \rightarrow \infty $, resulting the well
known decreasing $ \delta (r_{\text{min}})$ function out of the
horizon.) Rays passing both above and below $r_{\text{crit}}$ will
experience less deflection than $ \delta _{\text{max}}.$}
\label{fig4}
\end{figure}
\begin{figure}[tbp]
\includegraphics[height=7cm]{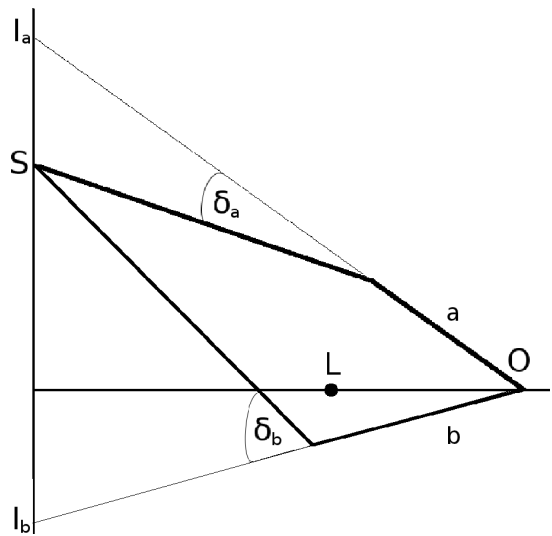}
\caption{The lensing geometry and rays labeled (a) and (b) for a large $%
 \omega $, reproducing image formation in Schwarzschild space-time,
with two images I$_{a}$ and I$_{b}$ of the source S. The deflection angles
obey $ \delta _{\left( b\right) }> \delta _{\left( a\right) }$%
. }
\label{fig5}
\end{figure}
\begin{figure}[tbp]
\includegraphics[width=18cm]{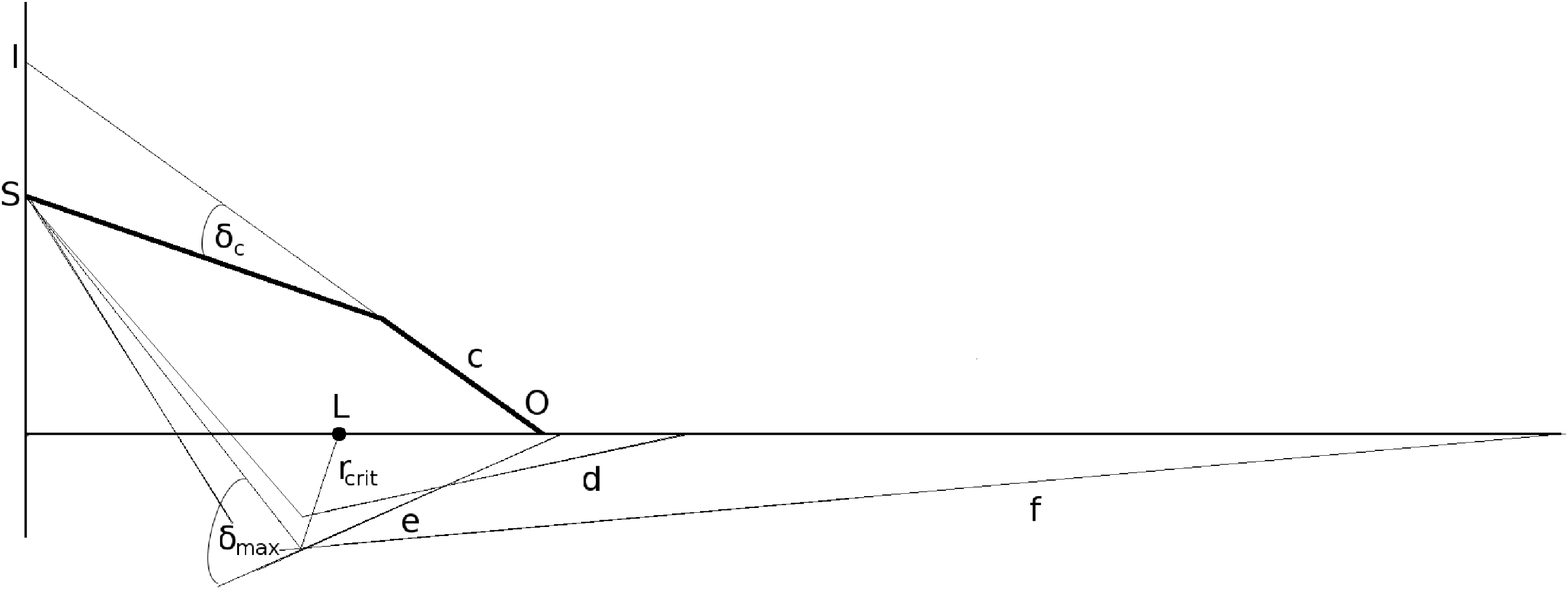}
\caption{The lensing geometry and rays for an $ \omega $ producing a
small enough value of $ \delta _{\text{max}}$, such as the upper
image (c) is formed, however no lower image appears, as the ray (e) passsing
through $r_{\text{crit}}$ will intersect the optical axis farther then the
observer. All rays passing below the lens, irrespectively whether their $r_{%
\text{min}}$ is smaller (d) or larger (f) then $r_{\text{crit}}$ will
intersect the optical axis at even larger distance.}
\label{fig6}
\end{figure}
For generic positions $\beta >0$, the numerical solving code for Eqs.~ (\ref%
{deltax0}), (\ref{rela}) and (\ref{lens_Ellis}) calculates $\left( \theta
/\beta ,r_{\text{min}}/d,\delta \right) $ as a function of the parameters $%
\bar{\lambda}$ and $\bar{\varepsilon}$. The image positions (in units $\beta 
$, thus represented as $\theta /\beta $) for a range of masses (represented
as $\bar{\varepsilon}/\beta ^{2}$) are plotted on Fig. \ref{fig3} (color
online). The three surfaces on the figure refer to the following situations:
(1) the focused positive image is represented by the upper surface (red in
color version); (2) the focused negative image is seen underneath (blue);
(3) the scattered images (with $0<\theta <\beta $) are found in the junction
of the two surfaces mentioned generated by a negative mass (green). There is
a parameter region encompassing only negative masses, where no image is
formed at all (see (\ref{cond})).

The parabolic edge of the represented surface (the parabola with the largest
opening given by the $\bar{\lambda}=0$ section) corresponds to image
formation by weak lensing in the Schwarzschild limit of the Kehagias-Sfetsos
space-time. As we increase $\bar{\lambda}$, a similar evolution of the loci
of the positive focused images occurs, as for the radius of the Einstein
ring: the images come closer to the optical axis. Increasing $\bar{\lambda}$
even more leads to a surprising situation: the negative image suddenly
disappears (the\ lower surface has a sharp edge, when $\bar{\lambda}%
\rightarrow \mathcal{O}\left( 10^{-3}\right) $). We will explain this
phenomenon below. Finally, for sufficiently large $\bar{\lambda}$ the
space-time flattens and $\theta /\beta \rightarrow 1$, corresponding to no
deflection at all.

In order to explain what happens in Fig \ref{fig3} at intermediate values of 
$\bar{\lambda}$ first we have to study the behavior of the deflection angle $%
\delta $ as function of the distance of closest approach $r_\text{min}$.
This is depicted in Fig \ref{fig4} for various parameter values $\omega $
corresponding to the intermediate $\bar{\lambda}$ range. We find that (a) at
a certain $r_\text{min}$ the deflection angle decreases with $\omega $, and
(b) there is a maximal deflection angle $\delta _\text{max}$, corresponding
to certain $r_\text{crit}$. Rays passing both above and below $r_\text{crit}$
will experience less deflection, than the one passing through $r_\text{crit}$%
.

Rays captured by the observer which pass below the lens should exhibit a
larger deflection angle, than the ones passing above the lens, see Fig \ref%
{fig5}. However for each source-lens-observer geometry there will be a value
of $\omega $ for which the corresponding $\delta _\text{max}$ will not be
sufficient to deflect any of the rays passing below the lens to the
observer. Hence the lower image disappears. This feature is illustrated in
Fig \ref{fig6}. Similar considerations hold for the negative mass region.

We conclude that light deflection by weak gravitational lensing typically
produces two images, as in Schwarzschild lensing, if $\omega $ is very
large. For any fixed source-lens-observer geometry there is an intermediate
value range of $\omega $-s for which only the upper image exists. Finally,
as we have already emphasized, a sufficiently large $\bar{\lambda}$ renders
the space-time close to flat.

\section{Einstein angles and the Kehagias-Sfetsos parameter \label{secV}}

\begin{table*}[t]
\begin{tabular}{|c|c|c|c|c|c|c|c|}
\hline
galaxy & $\theta _{E}$ [arcsec] & M [10$^{10}$ $M_{\odot }$] & $D_{L}$ [Mpc]
& $D_{S}$ [Mpc] & $r_\text{min}$ [Kpc] & $\omega _\text{min}$ $\left[
10^{-48}\text{ cm}^{-2}\right] $ & $\omega _\text{0,min}$ $[10^{-16}]$ \\ 
\hline
J0008-0004 & 1.16 & 35 & 1172.743 & 1708.956 & 6.595 & 0.18942 & 5.0657 \\ 
J0029-0055 & 0.96 & 12 & 750.391 & 1622.253 & 3.492 & 0.27033 & 0.84982 \\ 
J0037-0942 & 1.53 & 29 & 669.889 & 1411.834 & 4.969 & 0.34628 & 6.3576 \\ 
J0044+0113 & 0.79 & 9 & 446.200 & 672.576 & 1.709 & 1.2219 & 2.1606 \\ 
J0109+1500 & 0.69 & 13 & 905.964 & 1291.699 & 3.031 & 0.33425 & 1.2332 \\ 
J0157-0056 & 0.79 & 26 & 1276.277 & 1618.923 & 4.888 & 0.23749 & 3.5047 \\ 
J0216-0813 & 1.16 & 49 & 984.161 & 1289.158 & 5.535 & 0.35670 & 18.697 \\ 
J0252+0039 & 1.04 & 18 & 875.400 & 1644.710 & 4.414 & 0.22816 & 1.6138 \\ 
J0330-0020 & 1.10 & 25 & 1020.790 & 1676.984 & 5.444 & 0.20053 & 2.7361 \\ 
J0405-0455 & 0.80 & 3 & 293.680 & 1555.222 & 1.139 & 1.1703 & 0.22993 \\ 
J0728+3835 & 1.25 & 20 & 696.489 & 1463.700 & 4.221 & 0.32141 & 2.8066 \\ 
J0737+3216 & 1.00 & 29 & 964.236 & 1358.283 & 4.675 & 0.30324 & 5.5673 \\ 
J0822+2652 & 1.17 & 24 & 784.904 & 1372.544 & 4.452 & 0.30991 & 3.8969 \\ 
J0903+4116 & 1.29 & 45 & 1157.167 & 1675.061 & 7.237 & 0.19736 & 8.7248 \\ 
J0912+0029 & 1.63 & 40 & 580.506 & 968.258 & 4.587 & 0.60497 & 21.131 \\ 
J0935-0003 & 0.87 & 41 & 1013.21 & 1213.066 & 4.274 & 0.48386 & 17.756 \\ 
J0936+0913 & 1.09 & 15 & 653.631 & 1366.016 & 3.454 & 0.36699 & 1.8026 \\ 
J0946+1006 & 1.38 & 29 & 737.793 & 1388.471 & 4.936 & 0.32025 & 5.87966 \\ 
J0956+5100 & 1.33 & 37 & 782.476 & 1217.394 & 5.045 & 0.37351 & 11.163 \\ 
J0959+4416 & 0.96 & 17 & 775.156 & 1299.251 & 3.608 & 0.33752 & 2.1294 \\ 
J0959+0410 & 0.99 & 8 & 465.339 & 1304.229 & 2.233 & 0.58732 & 0.82058 \\ 
J1016+3859 & 1.09 & 15 & 592.045 & 1171.202 & 3.129 & 0.47174 & 2.3171 \\ 
J1020+1122 & 1.20 & 34 & 879.827 & 1326.068 & 5.119 & 0.31376 & 7.9181 \\ 
J1023+4230 & 1.41 & 23 & 656.356 & 1470.562 & 4.487 & 0.34248 & 3.9551 \\ 
J1029+0420 & 1.01 & 6 & 393.820 & 1394.687 & 1.928 & 0.73490 & 0.57756 \\ 
J1100+5329 & 1.52 & 47 & 954.102 & 1584.325 & 7.031 & 0.22552 & 10.875 \\ 
J1106+5228 & 1.23 & 9 & 366.879 & 1119.963 & 2.188 & 0.90303 & 1.5968 \\ 
J1112+0826 & 1.49 & 45 & 859.747 & 1408.858 & 6.211 & 0.28347 & 12.531 \\ 
J1134+6027 & 1.10 & 13 & 548.213 & 1223.120 & 2.924 & 0.49304 & 1.8190 \\ 
J1142+1001 & 0.98 & 17 & 737.793 & 1264.429 & 3.505 & 0.36081 & 2.2764 \\ 
J1143-0144 & 1.68 & 19 & 400.476 & 1111.614 & 3.262 & 0.79571 & 6.2708 \\ 
J1153+4612 & 1.05 & 11 & 626.027 & 1593.810 & 3.187 & 0.34366 & 0.90778 \\ 
J1204+0358 & 1.31 & 17 & 580.506 & 1410.844 & 3.687 & 0.41199 & 2.5993 \\ 
J1205+4910 & 1.22 & 25 & 719.910 & 1233.014 & 4.258 & 0.37895 & 5.1705 \\ 
J1213+6708 & 1.42 & 14 & 455.803 & 1419.669 & 3.138 & 0.7912 & 2.4779 \\ 
J1218+0830 & 1.45 & 16 & 493.548 & 1487.965 & 3.470 & 0.5017 & 2.8041 \\ 
J1250+0523 & 1.13 & 18 & 762.845 & 1545.386 & 4.179 & 0.27744 & 1.9624 \\ 
J1402+6321 & 1.35 & 29 & 693.857 & 1233.014 & 4.541 & 0.38808 & 7.1250 \\ 
\hline
\end{tabular}%
\caption{Column 1: the lens galaxies. Column 2-5: the Einstein angle $%
 \theta _{E}$, total lens mass inside the Einstein radius $R_{E}$,
the distances $D_{L}$ and $~D_{LS}$. The quantities calculated from the
model are $r_\text{min}$ (column 6), $ \omega _\text{min}$ and $%
 \omega _\text{0,min}:=G^{2}M_{lum+dark}^{2} \omega _\text{min%
}/c^{4}$ (columns 7 and 8).}
\label{table1a}
\end{table*}

\begin{table*}[t]
\begin{tabular}{|c|c|c|c|c|c|c|c|}
\hline
galaxy & $\theta _{E}$ [arcsec] & M [10$^{10}$ $M_{\odot }$] & $D_{L}$ [Mpc]
& $D_{S}$ [Mpc] & $r_\text{min}$ [Kpc] & $\omega _\text{min}$ $\left[
10^{-48}\text{ cm}^{-2}\right] $ & $\omega _\text{0,min}$ $[10^{-16}]$ \\ 
\hline
J1403+0006 & 0.83 & 10 & 650.899 & 1221.693 & 2.619 & 0.41355 & 0.90280 \\ 
J1416+5136 & 1.37 & 37 & 916.645 & 1555.865 & 6.088 & 0.23665 & 7.0726 \\ 
J1420+6019 & 1.04 & 4 & 250.179 & 1304.222 & 1.261 & 1.6168 & 0.56474 \\ 
J1430+4105 & 1.52 & 54 & 886.429 & 1351.555 & 6.532 & 0.30198 & 19.224 \\ 
J1436-0000 & 1.12 & 23 & 886.429 & 1551.984 & 4.813 & 0.24263 & 2.8020 \\ 
J1443+0304 & 0.81 & 6 & 490.443 & 1139.617 & 1.926 & 0.59752 & 0.46959 \\ 
J1451-0239 & 1.04 & 8 & 462.168 & 1285.325 & 2.330 & 0.59786 & 0.83530 \\ 
J1525+3327 & 1.31 & 48 & 1033.889 & 1487.965 & 6.566 & 0.25047 & 12.598 \\ 
J1531-0105 & 1.71 & 27 & 568.859 & 1509.087 & 4.716 & 0.40490 & 6.4438 \\ 
J1538+5817 & 1.00 & 9 & 518.128 & 1299.251 & 2.512 & 0.50680 & 0.89616 \\ 
J1621+3931 & 1.29 & 29 & 794.563 & 1381.108 & 4.969 & 0.30476 & 5.5953 \\ 
J1627-0053 & 1.23 & 23 & 701.736 & 1290.430 & 4.185 & 0.36381 & 4.2014 \\ 
J1630+4520 & 1.78 & 49 & 801.749 & 1544.046 & 6.919 & 0.26412 & 13.844 \\ 
J1636+4707 & 1.09 & 18 & 752.893 & 1451.370 & 3.979 & 0.29974 & 2.1201 \\ 
J2238-0754 & 1.27 & 13 & 499.736 & 1484.718 & 3.077 & 0.49325 & 1.8198 \\ 
J2300+0022 & 1.24 & 30 & 755.390 & 1208.707 & 4.541 & 0.38187 & 7.5028 \\ 
J2303+1422 & 1.62 & 27 & 554.146 & 1281.465 & 4.352 & 0.46848 & 7.4556 \\ 
J2321-0939 & 1.60 & 12 & 318.486 & 1300.500 & 2.470 & 1.0661 & 3.3514 \\ 
J2341+0000 & 1.44 & 22 & 642.666 & 1553.284 & 4.487 & 0.33734 & 3.5644 \\ 
\hline
\end{tabular}%
\caption{Table~ \ref{table1a} continued.}
\label{table1b}
\end{table*}

The Large Synoptic Survey Telescope (LSST) is expected to discover a high
number of gravitational lenses, allowing statistical studies \cite{LSSTbook}%
. An advantage of LSST will be its excellent image quality \cite{LSSTbook}.
The high resolution is crucial for lens searches, as the typical angular
scales of lensing are comparable to the seeing sizes of ground-based
observations (Fig 12.3 in \cite{LSSTbook}). The lens galaxy population is
expected \cite{LSSTbook} to be dominated by massive elliptical galaxies at
redshift 0.5-1, whose background light sources are the faint blue galaxies.
Therefore the detection of such systems depends on the ability to
distinguish lens light from the source light \cite{LSSTbook}. Fig 12.6 in 
\cite{LSSTbook} shows how the detection rate of galaxy lenses depends on
image quality. The best predicted seeing presented in the figure is 0.3
arcsec. According to Fig 12.5 in \cite{LSSTbook} the median seeing is about
half of that of the SDSS (Sloan Digital Sky Survey), which is 1.4 arcsec.

The Sloan Lens Advanced Camera and Spectrograph Survey (SLACS) has provided
the largest sample of galaxy-scale lenses to date, with almost 100 lenses
detected and measured \cite{Bolton}. The sources are faint blue galaxies,
selected by their emission lines appearing in the lower redshift SDSS
luminous red galaxy spectra. The largest collection of gravitational lens
systems of SLACS was analyzed in \cite{GGLP}. SLACS is a project that
combines the massive data volume of the SDSS with the high-resolution
imaging capability of the Hubble Space Telescope (HST) to identify and study
a large and uniform sample of gravitational lens galaxies. The lens galaxies
are selected from the spectroscopic database of the SDSS for the presence of
two galaxies along the same line of sight in the sky, one much more distant
than the other.

The HST images allow us to measure the angular size of the Einstein rings.
The observed Einstein angles are given in Table 1 in \cite{GGLP}, they range
from 0.69 arcsec to 1.78 arcsec. These angles combining with the distances
measured from the SDSS spectra give us direct measurements of the enclosed
masses of the nearer galaxies (the lens galaxies). The ACS-WFC (Advanced
Camera for Surveys, Wide Field Channel) instrument was used for measurement
of the Einstein angles \cite{Bolton}. This channel will be optimized for
surveys in the near-infrared to search for galaxies and clusters in the
early universe, and possesses 0.049 arcsec pixel size \cite{acswfc}. The
SLACS sample is currently the largest collection of gravitational lens
systems with known distances (redshifts) to both components.

The photometric and spectroscopic measurements for the 57 massive early-type
lens galaxies discussed in \cite{GGLP} are available from SDSS. By using
multicolor photometry and lens models, \cite{GGLP} studied stellar-mass
properties and the luminous and dark matter composition of the early-type
lens galaxies. The fraction of mass in the form of stars of the selected
early-type grade-A lens galaxies in the sample was also presented.

Astronomical angle measurements come with certain uncertainty $\Delta \theta 
$, which allow for a range $\omega \in \left( \omega _\text{min},\infty
\right) $ of the Ho\v{r}ava-Lifshitz parameter, where $\omega _\text{min}$
obeys $\theta _{E}(\omega _\text{min})=\theta _{E,Sch}-\Delta \theta $. For
any such $\omega $ the Einstein angle would be observationally
indistinguishable from the Schwarzschild value $\theta _{E,Sch}$. By
contrast, every $\omega <\omega _\text{min}$ correspond to Einstein rings
outside the measurement accuracy. In the following we take the accuracy of
SDSS, $\Delta \theta =0.049$ arcsec.

In Tables~\ref{table1a} and \ref{table1b} column 2 gives the Einstein angles
of the lens galaxies enlisted in column 1, with the total (dark + luminous)
masses falling inside the effective Einstein radius\textit{\ }$%
R_{E}=D_{L}\theta _{E}$ given in column 3 \cite{GGLP}. We have converted the
redshifts $z_{L}$, $z_{S}$ of the lens and the source galaxies given in
Table~1 in \cite{GGLP} to the angular diameter distances $D_{L}$, $D_{LS}$
(columns 4, 5) using the value $H_{0}=70$ $\mathrm{km/s/Mpc}$ for the Hubble
parameter and $\Omega _{\Lambda }=0.7$, $\Omega _{M}=0.3$ for the
cosmological parameters.

We confronted all these data with the weak lensing equations in the
Kehagias-Sfetsos space-time, obtaining numerically the values of $r_{\text{%
min}}$, $\omega _{\text{min}}$ and $\omega _\text{0,min}:=\left(
GM_{lum+dark}/c^{2}\right) ^{2}\omega _\text{min}$, given in columns 6, 7,
8. We will compare these values with other constraints existing in the
literature in the Conclusion Section, basically finding agreement with other
type of constraints.

\section{Strong lensing in the Kehagias-Sfetsos space-time \label{secVI}}

\begin{table*}[t]
\caption{Radii of the first and second relativistic Einstein rings, with
corresponding $ \omega _\text{min}$ and $ \omega _\text{0,min}$
for $\Delta \theta _{E}=10^{-5}$ arcsec. }
\label{table2}%
\begin{tabular}{|c|c|c|c|}
\hline
Einstein ring & $\theta _{E},_{Sch}$ $[10^{-5}$ arcsec$]$ & $\omega _\text{%
min}$ $[cm^{-2}]$ & $\omega _\text{0,min}$ \\ \hline
$1^{\text{st}}$ relativistic & $2.557$ & $8.1315\cdot 10^{-25}$ & $0.3282$
\\ 
$2^{\text{nd}}$ relativistic & $2.554$ & $8.1315\cdot 10^{-25}$ & $0.3282$
\\ \hline
\end{tabular}%
\end{table*}

Relativistic images are not observed yet, however the Multi-AO Imaging
Camera for Deep Observations (MICADO) - to function from 2018, using
adaptive optics, on the 42 m European Extremely Large Telescope - is
designed to have a resolution in the astrometric mode of about $10^{-5}$
arcsec \cite{binnun3,micado}. This resolution is close to the scale of the
relativistic images.

This will allow to constrain the quantities $\omega $ and $\omega _{0}$ in
the strong lensing configuration, suggested for testing the Schwarzschild
space-time in Ref. \cite{Bin-Nun}. In this configuration the lens is the
Supermassive Black Hole in the center of our galaxy, Sagittarius A*, with
mass $4.3\times 10^{6}$ $M_{\odot }$, and the light is coming from a stellar
source on the opposite side of the galaxy, such that the distances of the
lens and the source are $D_{L}=8.3$ kpc and $D_{S}=2D_{L}$. Next we apply
the method of constraining $\omega _{0}$ developed in the previous section
for this configuration and the expected resolution of MICADO.

The results are presented in Table~\ref{table2}. The rows refer to the first
and second relativistic Einstein rings, respectively, as calculated in the
Schwarzschild space-time (column 2). Columns 3 and 4 contain the lower limit
of $\omega $ and $\omega _{0}$ arising from the envisaged resolution of $%
\Delta \theta =10^{-5}$ arcsec. Although several orders of magnitude larger
than presently available limits, this value of $\omega _\text{0,min}$ would
still allow for the galactic Supermassive Black Hole to be a naked
singularity, as the black hole conditions $\omega _{0}>1/2$ and (\ref{BHcond}%
) are disobeyed. A further slight decrease in $\Delta \theta $ would be
necessary in order to disrule this possibility.

\section{Conclusion \label{sec:concl}}

In this paper, we discussed the image formation in both weak and strong
gravitational lensing by the Kehagias-Sfetsos solution in the framework of Ho%
\v{r}ava-Lifshitz gravity. Such a gravitational lens is characterized by a
mass-type parameter $\bar{\varepsilon}$ and an additional parameter $\bar{%
\lambda}$. The overbar denotes certain scaling of these dimensionless
parameters with a characteristic length $d$ characterizing the geometry. The
Schwarzschild limit occurs when $\bar{\lambda}\rightarrow 0$, while in the
limit $\bar{\lambda}\rightarrow \infty $ the Kehagias-Sfetsos space-time
becomes flat (the contribution of an increasing $\bar{\lambda}$ cancels the
gravitational attraction of the positive $\bar{\varepsilon}$). There is also
an intermediate range, characterized by the existence of a maximal
deflection angle $\delta _\text{max}$, occurring at $r_\text{crit}$. This
means that both the rays passing closer and farther to the lens than $r_%
\text{crit}$ will experience less deflection.

In the weak lensing approximation a closed system of equations for the
variables $\theta $ (image position), $r_\text{min}$ (distance of minimal
approach) and $\delta $ (deflection angle) is given by the lens equation
relating $\theta $ and $\delta $ to each other, an integral formula to
produce $\delta $ as a function of $r_\text{min}$ and a third equation which
connects $r_\text{min}$ to the impact parameter $b=D_{L}\sin (s\theta )$. We
have employed the Virbhadra-Ellis lens equation, presenting sufficient
accuracy for our first order approach. For the deflection\ angle we adopted
the improper integral given in Ref. \cite{WEINBERG}, then performed a
transformation which removes the singularity at $r=r_\text{min}$, allowing
the numerical integration. Finally we employed the algebraic relation (\ref%
{rela}) connecting $\theta $ and $r_\text{min}$ given in Ref. \cite{GERGELY}.

The aligned case $\beta =0$ leads to the formation of Einstein rings. These
Einstein angles were plotted as functions of the space-time parameters $\bar{%
\varepsilon}$ and $\bar{\lambda}$ in Fig \ref{fig2}. For $\bar{\lambda}%
\rightarrow 0$ we recovered the half-parabola shaped section of the surface,
representing the Schwarzschild limit of weak lensing. As $\bar{\lambda}$
increases, the opening of the half-parabola decreases, indicating a
weakening of gravity.

For source positions $\beta >0$ the surface representing the image
positions, shown in Fig \ref{fig3}a, is more complex. We showed that in the
weak gravitational lensing regime, the light deflection either produces two
images, as in Schwarzschild lensing, or only one image. In the latter case
the existence of $\delta _{\text{max}}$ is obstructing the creation of the
lower image. As $\bar{\lambda}$ increases further, the space-time flattens
and $\theta /\beta \rightarrow 1,$ which causes that the $\theta >0$ (upper,
red) surface on Fig \ref{fig3}a flattens as well. When $\bar{\lambda}\rightarrow 
\mathcal{O}\left( 10^{-3}\right) $ the $\theta <0$ (lower, blue) branch has
a sharp edge and disappears. As a further consequence the middle (green)
surface representing the two scattered images occurring for negative masses
also disappear for $\bar{\lambda}\rightarrow \infty $. As in the
Schwarzschild case, there is a parameter region encompassing only negative
masses, where no image is formed at all.

We also analyzed the photometric and spectroscopic measurements for a sample
of 57 lens galaxies available from SDSS, given in Ref. \cite{GGLP}. These
led to estimates of masses and lensing distances. From the observed
locations of the corresponding Einstein rings and the accuracy of
measurements we derived the range of the Ho\v{r}ava-Lifshitz parameter $%
\omega =\bar{\lambda}^{-1}d^{-2}$ characterizing the Kehagias-Sfetsos
space-time compatible with the observations. The results are presented in
Tables \ref{table1a}, \ref{table1b} . The dimensionless quantity $\omega _%
\text{0,min}:=\left( GM_{lum+dark}/c^{2}\right) ^{2}\omega _\text{min}$ for
the sample of lens galaxies is typically found to be of the order $10^{-16}$.

We compare these numbers with related results in the literature. In Ref. 
\cite{Harko:2009qr} the Solar System tests were analyzed, imposing
constraints on $\omega _{0}$ from the available observations. Perihelion
precession of the planet Mercury, deflection of light by the Sun and the
radar echo delay gave, respectively the limits $\omega _{\text{0,min}%
}^{(pp)}=6.9\times 10^{-16}$ ,$\omega _{\text{0,min}}^{(ld)}=1.1\times
10^{-15}$, and $\omega _{\text{0,min}}^{(red)}=2\times 10^{-15}$,
respectively. In Ref. \cite{ir1} the weak-field and slow-motion
approximation was employed to compare with the orbital periods of the
transiting extrasolar planet HD209458b (Osiris), obtaining a weaker bound of 
$\omega _{0}^{(Osiris)}=1.4\times 10^{-18}$. In Ref. \cite{Liu}, constraints
on Ho\v{r}ava-Lifshitz gravity from light deflection observations including
long-baseline radio interferometry, Jupiter measurement, Hipparcos satellite
are found also in the range of $\omega _{0}^{\left( radio\right) }\in
10^{-15}\div 10^{-17}$. Our findings are therefore consistent with these
results. Stronger bounds of $\omega _{o}^{\left( residual\right) }=7.2\times
10^{-10}$ were presented from the accurate data of range-residuals of the
planet Mercury ranged from the Earth, and $\omega _{0}^{\left( Sag\right)
}=8\times 10^{-10}$ for the system constituted by the S2 star orbiting the
Supermassive Black Hole (Sagittarius A*) in the center of the Galaxy \cite%
{ir2}. We stress however, that the latter value of $\omega _{0}^{\left(
Sag\right) }$ would render our galactic Supermassive Black Hole into a naked
singularity, as the condition $\omega _{0}>1/2$ is disobeyed.

Finally, we discussed the first two relativistic Einstein rings in the
strong lensing regime. Applying to the galactic center as a strong lens and
a light source located on the opposite side, a configuration discussed in
Ref. \cite{Bin-Nun}, we determined the constraints on $\omega _{0}$, given
in Table \ref{table2} (under the assumption of fixed lens mass) arising from
the expected accuracy of $10^{-5}$ arcsec of future instruments \cite{micado}%
. We found that such measurements would constrain quite severely the
parameter range, up to $\omega _\text{0,min}$ of order $10^{-1}$, allowing
to either falsify the Ho\v{r}ava-Lifshitz theory or to render the parameter
of the Kehagias-Sfetsos space-time into a regime where it practically
becomes indistinguishable from the Schwarzschild space-time. This would set
the strongest observational constraint on the Ho\v{r}ava-Lifshitz parameter
up to date.

\section{Acknowledgments}

L. \'{A}. G. and F. S. N. L. would like to thank the Department of Physics
of The University of Hong Kong for their support and warm hospitality during
the preparation of this work. Zs. H. was supported by the European Union and
co-funded by the European Social Fund through the T\'{A}MOP
4.2.2/B-10/1-2010-0012 grant. L. \'{A}. G. was partially supported by COST
Action MP0905 \textquotedblleft Black Holes in a Violent
Universe\textquotedblright . The work of T. H. was supported by the General
Research Fund grant number HKU 701808P of the government of the Hong Kong
Special Administrative Region. F. S. N. L. acknowledges financial support of
the Funda\c{c}\~{a}o para a Ci\^{e}ncia e Tecnologia through the grants
PTDC/FIS/102742/2008, CERN/FP/109381/2009 and CERN/FP/116398/2010.

\end{document}